# Probing length-scale separation of thermal and spin currents by nanostructuring YIG

*Asuka Miura[1], Takashi Kikkawa[2,3], Ryo Iguchi[2], Ken-ichi Uchida[2,4,5,6]\*, Eiji Saitoh[2,3,6,7], Junichiro Shiomi [1,8,\*]*

[1] Department of Mechanical Engineering, The University of Tokyo, Tokyo 113-8656, Japan

[2] Institute for Materials Research, Tohoku University, Sendai 980-8577, Japan

[3] WPI Advanced Institute for Materials Research, Tohoku University, Sendai 980-8577, Japan

[4] National Institute for Materials Science, Tsukuba 305-0047, Japan

[5] PRESTO, Japan Science and Technology Agency, 4-1-8, Kawaguchi, Saitama 332-0012, Japan

[6] Center for Spintronics Research Network, Tohoku University, Sendai 980-8577, Japan

[7] Advanced Science Research Center, Japan Atomic Energy Agency, Tokai 319-1195, Japan

[8] CREST, Japan Science and Technology Agency, 4-1-8, Kawaguchi, Saitama 332-0012, Japan

**ABSTRACT**

We have fabricated bulk nanostructured ferrimagnetic materials with different grain sizes by sintering ball-milled $Y_3Fe_5O_{12}$ (YIG) nanoparticles and measured the grain-size dependence of the thermal conductivity and spin Seebeck thermopower. The nanostructuring reduces both thermal conductivity and thermopower, but the reduction of the latter was found to be considerably stronger despite the moderate difference in magnetization, which suggests that the length scales of transport of magnons and phonons contributing to the spin Seebeck effect are significantly larger than that of phonons carrying thermal current. This is consistent with the measurements of high-magnetic-field response of the spin Seebeck thermopower and low-temperature thermal conductivity, where the quenching of magnons seen in single-crystalline YIG was not observed in nanostructured YIG due to scattering of long-range low frequency magnons.

## I. INTRODUCTION

The spin Seebeck effect (SSE) is the phenomenon that spin current is generated from a temperature gradient in a magnetic material[1]. The spin current is injected from the magnet into an attached metal via the spin exchange interaction at the magnet/metal interface and can be detected as an electric voltage (SSE voltage) through the inverse spin Hall effect in the metal[2-7]. In particular, the longitudinal spin Seebeck effect (LSSE) configuration has been used in most of recent SSE studies because the LSSE system allows simpler SSE measurements [8]. Although spin Seebeck modules are promising for harvesting energy from ubiquitous heat sources, its efficiency is still limited compared to conventional Seebeck modules, and the enhancement of the figure-of-merit *ZT* is needed. Here, *ZT* is proportional to the square of the thermopower and inversely proportional to the thermal conductivity[9]. One approach for enhancing *ZT* is to reduce thermal conductivity of the magnetic layer by nanostructuring with a size smaller than the phonon mean free path to dominantly scatter phonons at the interfaces[10]. However, the key is to do so without sacrificing the thermoelectricity, therefore to avoid inhibition of magnon transport or phonon transport that drags the spin current in the magnetic layer. We note that the bulk magnon spin current induced by the temperature gradient across the thickness of the magnetic layer is essential for the existence of LSSE as shown by Rezende *et al*. in both the theory and experiment[11]. Therefore, to design the appropriate nanostructure, knowledge of the characteristic length-scales of thermal current and spin current in the magnetic layer is indispensable.

There have been some reports on the length scale of phonon and magnon transport in LSSE with $Y_3Fe_5O_{12}$ (YIG) as the magnetic material. Boona *et al*. investigated the effect of high magnetic field on temperature dependence of thermal conductivity and heat capacity in bulk single-crystalline YIG, and extracted the mean free path of phonons and magnons participating in heat conduction using a simple kinetic gas theory with gray approximation[12]. The obtained effective average mean-free-path of the thermal magnons is a few nanometers at room temperature, suggesting that LSSE is primarily driven by low-energy magnons with much larger mean free paths. They further suggested the minor role of the phonon-mediated processes, that is phonon drag[13], because magnon-phonon interaction with the low energies is expected to be weak[14]. The idea is supported by the investigation of the effect of film thickness in Pt/single-crystalline YIG systems[15-18], which shows that LSSE thermopower increased with film thickness beyond the mean free path of heat carrying magnons. However, recent research suggests the importance of the phonon-drag processes in the SSE by a simultaneous measurement of a LSSE thermopower and thermal conductivity in a Pt/single-crystalline YIG/Pt system[19]. Therefore, more systematic study of the correlation of thermal conductivity and LSSE thermopower is needed.

In this study, we prepared nanostructured YIG samples by sintering nanoparticles, which has been used as a conventional technique to fabricate nanostructured bulk thermoelectric materials[20,21], and investigated the temperature dependence of thermal conductivity and LSSE thermopower in the Pt/YIG systems. Nanostructuring reduces the thermal conductivity and thermopower by boundary scattering of phonons and magnons with a characteristic length scales of transport larger than the grain size, and therefore the characteristic length scales can be identified by comparing the properties of samples with various grain sizes (including single-crystal). We also reported the high-magnetic-field response of thermal conductivity and LSSE thermopower to confirm the contribution of low frequency magnons to heat conduction and spin current generation.

## II. EXPERIMENTAL METHOD

We prepared two batches of YIG nanoparticles with different average grain size from single crystal (Ferrisphere, Inc.) by planetary ball mill using premium line P-7 (Fritsch) firstly with stabilized zirconia balls of 10 mm in diameter for 2 hours, and then with those of 1 mm in diameter for 6 hours and 50 hours respectively at rotation speed of 300 rpm. The process was performed under the atmosphere. Each batch of the obtained YIG nanoparticles were sintered by plasma activated sintering (PAS) process using Ed-PAS IV (Elenix). In the process, the temperature is controlled by DC current and the resistance heating in combination with mechanical pressure shortens the sintering time. The YIG powders were loaded into a graphite die with diameter of 10 mm. After evacuating the PAS chamber, the temperature was raised to 1163 K and kept constant for 2 min with a pressure of 94 MPa. The density of the sintered samples was measured using the Alchimedes method at room temperature. The measured values of both two samples were 4.97 gcm$^{-3}$, 96.1% of the theoretical density of the single-crystalline YIG. Structural characterization was carried out using transmission electron microscopy (TEM), electron back scatter diffraction (EBSD), and electron probe microanalysis (EPMA).

Thermal conductivity of the four YIG samples (two nanostructured, single-crystalline, commercially produced polycrystalline (MITSUBISHI ELECTRIC METECS CO., LTD.) YIG) was measured by the steady-state method using PPMS (Quantum Design) in two-probe configuration from 2 K up to 300 K. Gold-plated copper leads and heat sinks were attached to both ends of the samples with silver epoxy paste, and two thermometers were used to determine the temperature difference imposed between both ends of the samples. The dimensions of samples are summarized in Table II. The measurements were carried out with and without a magnetic field of $H = 90$ kOe along the direction parallel to the temperature gradient.

LSSE thermopower of the single-crystalline, commercially-produced-polycrystalline, and nanostructured YIG samples was measured at temperatures ranging from 10 K up to 300 K by attaching a 10-nm-thick Pt film on the top and bottom surfaces of the YIG samples after the surfaces were mechanically polished with sand papers and alumina slurry[15,19,22]. Here, the dimensions of the samples are summarized in Table II. The Pt/YIG/Pt samples were sandwiched between two sapphire plates and a temperature difference of 10 K was imposed between the plates using a chip heater. The temperatures at the top and bottom of the samples was determined from the resistance of the Pt films[19,22]. The magnetic field of $H$ = 1.8 kOe was applied in the direction perpendicular to the temperature gradient. We also measured the magnetic-field dependence of the thermopower between ±90 kOe at 300 K using the Pt/YIG samples[15].

## III. RESULTS AND DISCUSSION

Measurement of the average grain size of commercially-produced-polycrystalline and nanostructured YIG samples was carried out using EBSD and the results are shown in Table II. The values of the average grain size decreased with increasing milling-time and agreed with the analysis using the low-resolution TEM (Fig. 1(a)). The high crystallinity is also evidenced by the diffraction-pattern image (the inset of Fig. 1(a)). Furthermore, the high-resolution TEM image clearly identified the interfaces between nanocrystals and there are few contaminants and precipitates at the grain boundaries (Fig. 1(b)). The EPMA analysis identified that the chemical composition of all the samples are $Y_3Fe_5O_{12}$ with small amount of impurities such as Pb and Mn, mixed in the process of growing the single-crystalline ingots. This confirms that there are few contaminants mixed with the powder during ball-milling. TEM images of the interfaces between Pt and YIG show that the surface roughness of all YIG samples after polishing is 0.2-0.3 nm and the morphologies are almost the same, which justifies the measurements of the grain-size effect on LSSE thermopower.

Figure 2(a) shows the temperature dependence of thermal conductivity of single-crystalline and polycrystalline samples including the two nanostructured YIG samples at $H = 0$ kOe and $H = 90$ kOe. The thermal conductivity of the polycrystalline samples at $H = 0$ kOe is significantly smaller than that of the single-crystalline sample over the whole temperature range. Furthermore, the values of thermal conductivity of the polycrystalline samples decrease with decreasing the average grain size. This is due to the increase in the scattering rate of the heat carrying phonons by grain boundaries. The lattice thermal conductivity in the phonon Boltzmann transport formalism is expressed as

$$\kappa = \frac{1}{3} \sum_s \int C_{v,s}(\omega) v_s^2(\omega) \tau_s(\omega) D_s(\omega) d\omega . \tag{1}$$

Here, $v$ is the phonon group velocity, $\tau$ is the phonon relaxation time, $D$ is the phonon density of states, and $C_v = \hbar\omega \partial f/\partial T$ is the phonon specific heat, where $\hbar$ is the reduced Planck constant, $\omega$ is the phonon frequency, $f$ is the Bose-Einstein distribution function, and $T$ is temperature. The relaxation time can be expressed by summing the rates of intrinsic phonon-phonon scattering, boundary scattering, and impurity scattering using Matthiessens's rule as

$$\frac{1}{\tau_s(\omega)} = \frac{1}{\tau_{ph,s}(\omega)} + \frac{v_s(\omega)}{L} + B\omega^4 , \tag{2}$$

where $\tau_{ph}$ is the relaxation time for phonon-phonon scattering, $L$ is the average grain size, and $B$ is a constant. Plugging Eq. (1) into Eq. (2) gives how nanostructuring (deceasing $L$) reduces thermal conductivity. The trend of average grain-size dependence of thermal conductivity agrees with the thickness dependence in thin film YIG previously reported[23]. However, the maximum thermal conductivity in the sample with average grain size of 0.55 µm, 5.2 Wm$^{-1}$K$^{-1}$, is much lower than that in 190 nm-thick film, 23 Wm$^{-1}$K$^{-1}$ [23], which indicates suitability of nanostructuring for reducing thermal conductivity. Note that the previous phonon transport calculations of silicon nanocrystalline structures have shown that the grain-size dependence of thermal conductivity is insensitive to the width of size distribution and determined by the average size[24].

We also found that the difference between the thermal conductivity below 10 K at $H = 0$ kOe and $H = 90$ kOe decreases with average grain size (Fig. 2(b)), resembling the trend observed for thin film YIG[23]. This is because magnons, the thermal excitation of which is suppressed by applying magnetic field, are scattered by grain boundaries and do not contribute to heat conduction in polycrystalline samples. The magnon density of states at low temperatures is given by[25]

$$D(\omega) = D_0 \sqrt{\hbar\omega - \mu_B g_L H} \,, \tag{3}$$

where $D_0$ is a constant, $\mu_B$ is the Bohr magneton, and $g_L$ is the Landé factor. Applying magnetic field creates a forbidden gap in the magnon band structure, and thus, reduces the density of states (The Zeeman gap $\mu_B g_L H$ at 90 kOe corresponds to ~0.25 THz in units of frequency). This leads to reducing magnon thermal conductivity of single-crystalline YIG as observed in the previous studies[12,26]. Note that the magnons suppressed by applying magnetic field have comparatively low frequency and large mean free path. Therefore, the boundary scattering of magnon with a large mean free path reduces the high-magnetic-field response of thermal conductivity. The response to magnetic field below 10 K is absent with average grain size of 2.25 μm but is present with average grain size of 20 μm. This indicates that the smallest mean free path of magnons with noticeable contribution to thermal conductivity below 10 K is between 2.25 μm and 20μm. This is consistent with the conclusion in previous work[12] that magnon thermal mean free path is more than about 10 μm at temperatures below 10 K.

Figure 2(c) and 2(d) shows the temperature dependence of LSSE thermopower in the single-crystalline and polycrystalline samples at temperatures ranging from 10 K to 300 K and at magnetic field of $H$ = 1.8 kOe. The thermopower of the polycrystalline samples is significantly smaller than that of the single-crystalline YIG in the whole temperature range and decreases with decreasing average grain size, which is the same trend as the average grain-size dependence of thermal conductivity. This reduction in the thermopower in polycrystalline YIG is, as observed by Uchida *et al*.[27], due to the decrease in the characteristic length scales of magnons and phonons contributing to generation of spin current due to boundary scattering. What exactly happens to spin current at the YIG grain boundaries remains unknown but the grain boundaries with few contaminants and precipitates, shown in Fig. 1, suggest that there is finite magnon transmission through the boundaries although the magnons should still be scattered. Although it has been observed that even sub-nanometer-thick nonmagnetic interlayer between Pt and YIG can terminate the spin current[28-30], the YIG grain boundaries in our samples have no such interlayers. Nevertheless, the results in Fig. 2(c) and 2(d) clearly show that nanostructuring reduces the relaxation time of magnons as it does for that of heat carrying phonons and thus leads to the decrease in the thermopower.

The reduction of the thermopower was found to be considerably stronger than that of thermal conductivity, which suggests that the length scales of transport of magnons and phonons contributing to the spin current are significantly larger than that of phonons carrying heat. This agrees with the above-discussed low-temperature measurements of thermal conductivity with and without magnetic field (Fig. 2(b)). We note that the magnitude of the magnetization of YIG is reduced by nanostructuring due to the nanosized dimensions of the grains as observed by Gaudisson *et al*.[31] and Sanchez *et al*.[32] and the saturation magnetization decreased with decreasing average grain size (Table III). However, the extent of the reduction is limited to about 10%, and thus this solely does not explain the reduction of the LSSE thermopower[33], and therefore supports that magnons scattering at grain boundaries plays the leading role in the reduction of the LSSE.

The peak temperature of thermal conductivity increased from 35 K to 127 K with decreasing average grain size (Fig. 3). The reduction of the average grain size enhances the boundary scattering of low frequency phonons, and thus, the effective frequency of heat carrying phonons (i.e. Debye temperature) increases. In addition, nanostructuring, by enhancing boundary scattering with respect to phonon-phonon scattering, weakens the negative temperature dependence in the high temperature regime. These result in a peak shift toward higher temperature. Note that, in YIG thin films with thicknesses ranging from 190 nm to 6.7 μm reported by Euler *et al*. [23], the peak temperature of out-of-plane thermal conductivity shifted from 57 K to 65 K. We also found that the peak temperature of the thermopower of polycrystalline samples increase with decreasing average grain size. The peak shift of thermopower is, similarly to that of thermal conductivity, due to the boundary scattering of low frequency magnons and phonons contributing to generation of spin current. We also found that the peak temperatures of thermopower is larger than those of thermal conductivity, which suggests that the characteristic length scales of transports of magnons and phonons contributing to the spin current are significantly larger than that of phonons carrying thermal current.

Figure 4 shows the high-magnetic-field response of the thermopower. The suppression of the LSSE seen in single-crystalline YIG was not observed in polycrystalline sample with average grain size of 2.25 μm. This is due to the increase in boundary scattering of low frequency magnons, which is consistent with high-magnetic-field response of the low-temperature thermal conductivity. Note that this also agrees with the previously reported film-thickness dependence of high-magnetic-field response of thermopower[15-18]

Figure 5 and Table III are shown to discuss the power-law scaling of the temperature dependence of thermal conductivity and thermopower of all the samples. The exponent $n$ is determined by fitting the measured data with a power-law function $CT^n$, where $C$ is the prefactor and $T$ is temperature. The absolute value of the exponent $|n|$ for thermal conductivity in temperature range above 100 K decreases with average grain size (Fig. 5(a)). At high temperatures, $C_v$ in Eq. (1) becomes Boltzmann constant $k_B$, and considering the fact $v$ and $D$ are nearly temperature-invariant, the temperature dependence of thermal conductivity is determined by that of $\tau$. In case of pure single crystal, as the relaxation is dominated by phonon-phonon scattering, thermal conductivity is inversely proportional to temperature. For single-crystalline YIG used in our study, $|n|$ is slightly smaller than 1 due to incorporation of small amount of impurities. The decrease of average grain size $L$ results in the reduction of temperature dependence of the relaxation time (Eq. (2)) and thus temperature dependence of thermal conductivity becomes weaker.

The exponent of the temperature dependence of thermal conductivity below 6 K with and without the magnetic field provides insight into phonon transmissivity through the grain boundary (Fig. 5(b)). In case of single crystal, $|n|$ is 2.28 when $H = 0$ kOe and it increases to 3 when $H = 90$ kOe. This makes sense because, at low temperature, the temperature dependence of magnon thermal conductivity scales with $|n| = 2$ [16], and that of phonon thermal conductivity scales with $|n| = 3$ for 3D materials. The combined value $|n| = 2.28$ is closer to that of magnon ($|n| = 2$) because the thermal conductivity of magnons is relatively larger than that of phonons at these temperatures as seen in Fig. 5(b). At these low temperatures, mean free paths of all the populated phonons are bounded by sample size (i.e. terminated at the sample surface), and thus, are temperature invariant. Therefore, temperature dependence of phonon thermal conductivity follows that of phonon specific heat (Eq. (1)). However, when forming grain boundaries, $|n|$ of phonon thermal conductivity ($H$ = 90 kOe) is significantly reduced as the average grain size decreases despite that the presence of grain boundary does not alter specific heat. This is because, unlike the sample surface, where the phonon transmission is zero, grain boundaries allow transmission of phonons whose probability in general is larger for lower frequency. In other words, mean free path (or relaxation time) decreases with increasing frequency. This counteracts the increase in specific heat of higher frequency modes when increasing temperature, and thus, weakens the temperature dependence of phonon thermal conductivity.

The effect of grain boundary on $|n|$ of thermopower is similar to that of thermal conductivity. At temperatures above 100 K, $|n|$ decreases with average grain size (Fig. 5(c)) due to the boundary scattering of magnons and phonons contributing to generation of spin current. At temperatures below 20 K, $|n|$ for the polycrystalline samples at $H$ = 1.8 kOe is slightly smaller than that for single crystal (Fig. 5(d)) and $|n|$ is about 1. This could be due to phonon and magnon transmissivity through the grain boundary. Note that the value corresponds to that for bulk YIG for the ballistic regime in the magnon Boltzmann transport formalism[16].

Figure 6 shows the relation between thermal conductivity and LSSE thermopower of all the samples in the whole temperature range. In single-crystalline YIG, there is a strong positive correlation between thermal conductivity and LSSE thermopower, consistent with the results in [19]. Importantly, we found that nanostructuring weakens the correlation. In particular, the relation between thermal conductivity and thermpower of the polycrystalline sample with average grain size of 20.0 μm is greatly different from that of single-crystalline YIG. This indicates that, in principle, thermal conductivity and LSSE thermpower can be optimized independently by using nanostructuring. Note that in this study the reduction of the thermopower was considerably stronger than that of thermal conductivity due to long-range magnon transport in YIG[34,35]. In future research our methodology will be applied to materials with a comparatively small length scale of spin currents and large length scale of thermal currents such as spinel ferrites ($CoFe_2O_4$[36], $NiFe_2O_4$[37]) to enhance *ZT*.

## IV. CONCLUSION

In summary, we fabricated nanostructured YIG samples with different grain sizes using plasma sintering of nanoparticles obtained from ball milling and measured the temperature dependence of thermal conductivity and LSSE thermopower. Both the thermal conductivity and LSSE thermopower are reduced compared to those of single crystal over the entire temperature range and the reduction of the latter is considerably stronger. This suggests that the length scales of transports of magnons and phonons contributing to the spin current are significantly larger than that of phonons carrying thermal current. This is consistent with the high-magnetic-field responses and the peak temperature shift of the thermopower and thermal conductivity. We also found that, in single-crystalline YIG, there is a strong positive correlation between thermal conductivity and thermopower, and the correlation is significantly weakened by nanostructuring. This indicates that thermal conductivity and thermpower can be controlled independently by using nanostructuring.

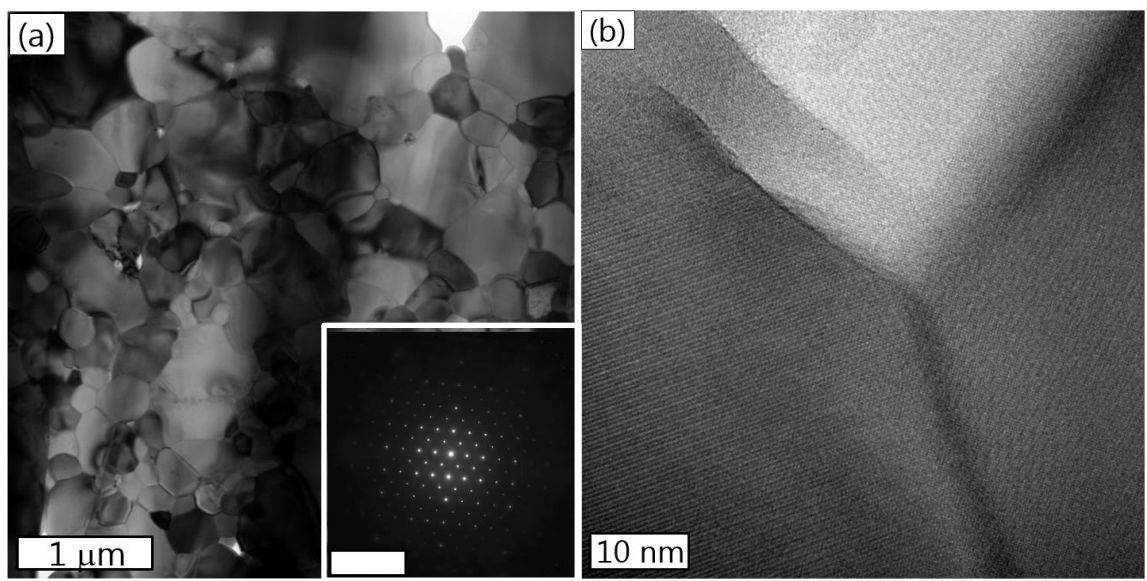

FIG. 1. (a) Low-resolution and (b) high-resolution TEM image of the nanostructured YIG with average grain size of 2.25 μm fabricated by sintering of the 6-hour-milled nanoparticles. The inset in (a) is the diffraction pattern and the scale bar represents 10 nm$^{-1}$.

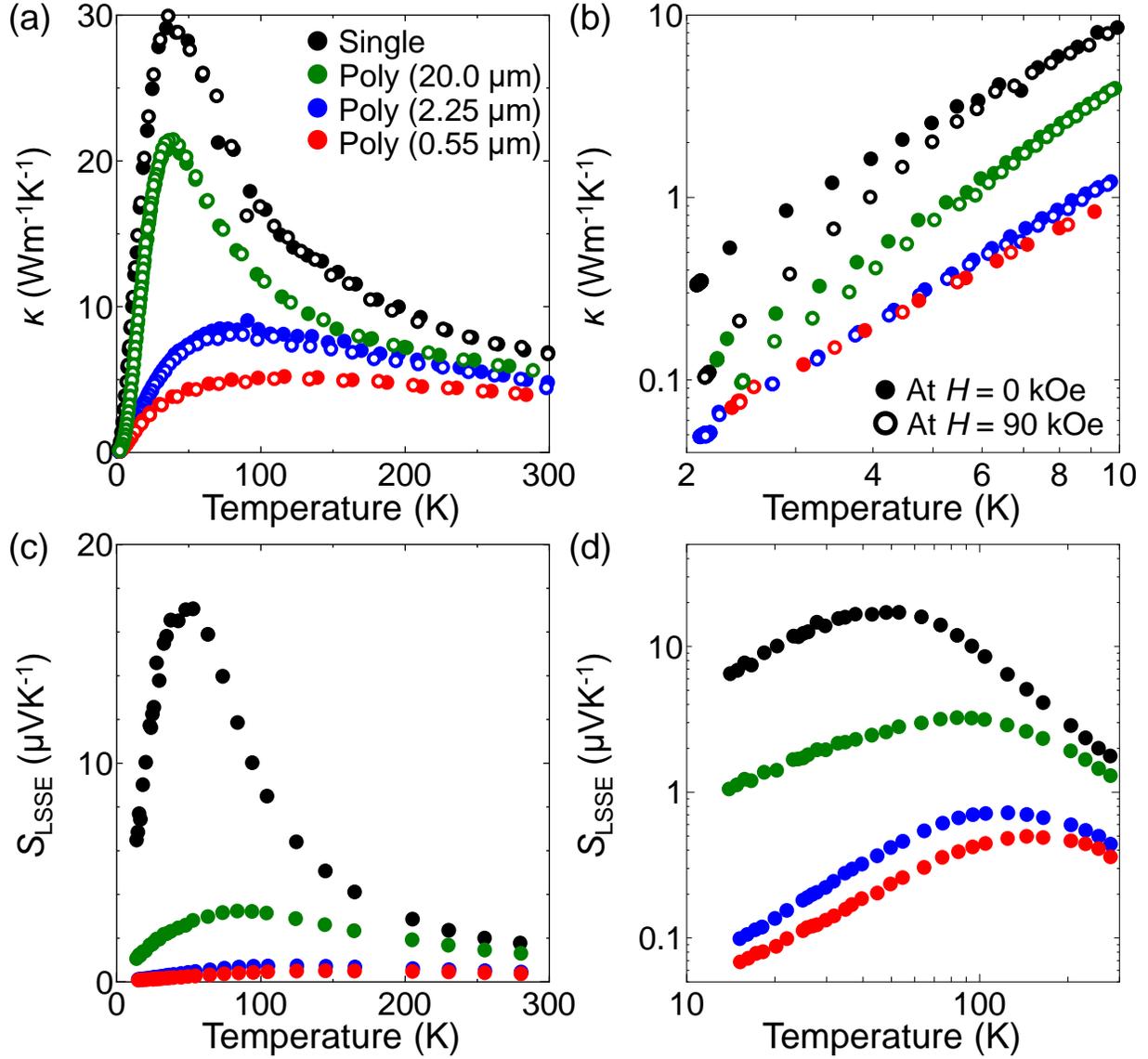

FIG. 2. (a) Comparison of temperature dendence of thermal conductivity in single-crystalline YIG (black) and polycrystalline samples with average grain size of 20.0 μm (green), 2.25 μm (blue) and 0.55 μm (red) at $H$ = 0 kOe (solid circle) and $H$ = 90 kOe (open circle) in a linear scale. (b) The enlarged view at temperature ranging from 2 K up to 10 K in a log-log scale. (c) Temperature dependence of LSSE thermopower at $H$ = 1.8 kOe in single-crystalline (black) and polycrystalline samples with average grain size of 20.0 μm (green), 2.25 μm (blue) and 0.55 μm (red) in a linear scale. (d) a log-log plot for (c).

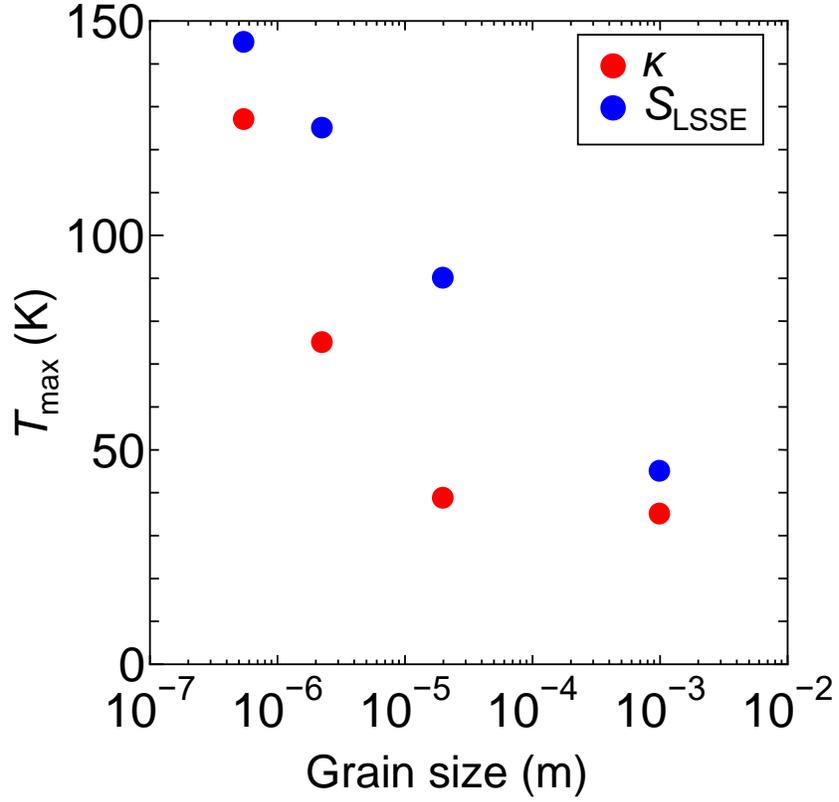

FIG. 3. Average grain-size dependence of peak temperature of thermal conductivity (red) and thermopower (blue). The peak temperature increased with decreasing average grain size due to the reduction of the characteristic length scales. The peak temperatures of thermopower are larger than those of thermal conductivity, which suggests that the characteristic length scales of transport of magnon and phonons contributing to the spin current are significantly larger than that of phonons carrying thermal current.

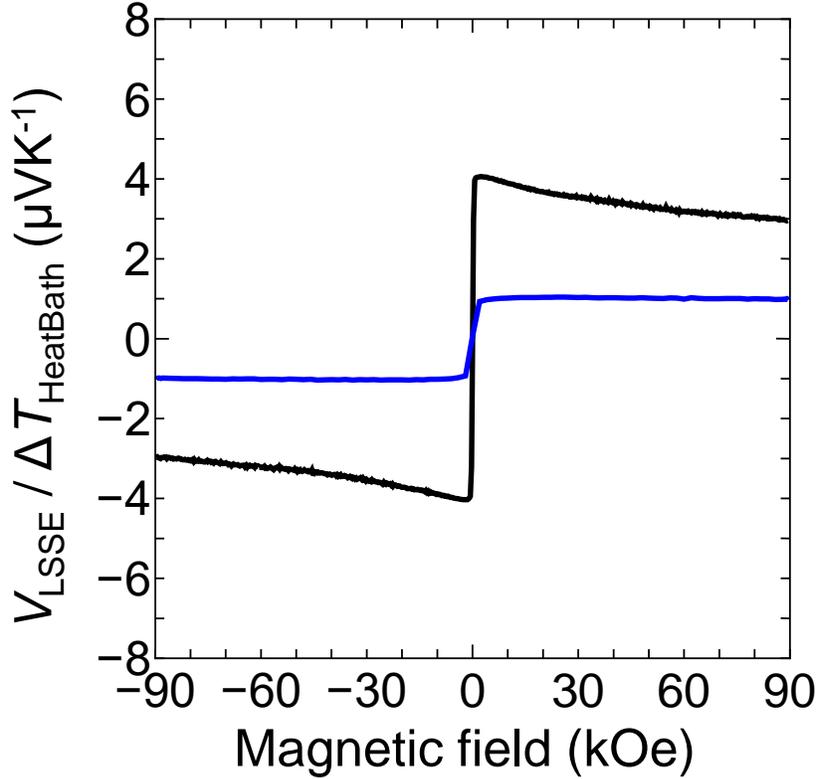

FIG. 4. $H$ dependence of $V_{LSSE}/\Delta T_{HeatBath}$ in single-crystalline YIG (black) and nanostructured YIG with average grain size of 2.25 μm (blue) between ±90 kOe at 300 K, where $V_{LSSE}$ is the LSSE voltage between the ends of the Pt layer and $\Delta T_{HeatBath}$ is the temperature difference between the top and bottom of the sample. Here $\Delta T_{HeatBath}$ is defined as the temperature difference between two heat baths. Note that the batches used in the measurement of $H$ dependence of $V_{LSSE}/\Delta T_{HeatBath}$ are different from those in the temperature dependence of LSSE thermopower (Fig. 2(c) and 2(d)) and the thickness of Pt attached on the samples is 5 nm. In the nanostructured sample, the reduction of LSSE thermopower on high magnetic field seen in single-crystalline YIG is not observed, which indicates that low frequency magnons contributing to the spin current are scattered by the grain boundaries.

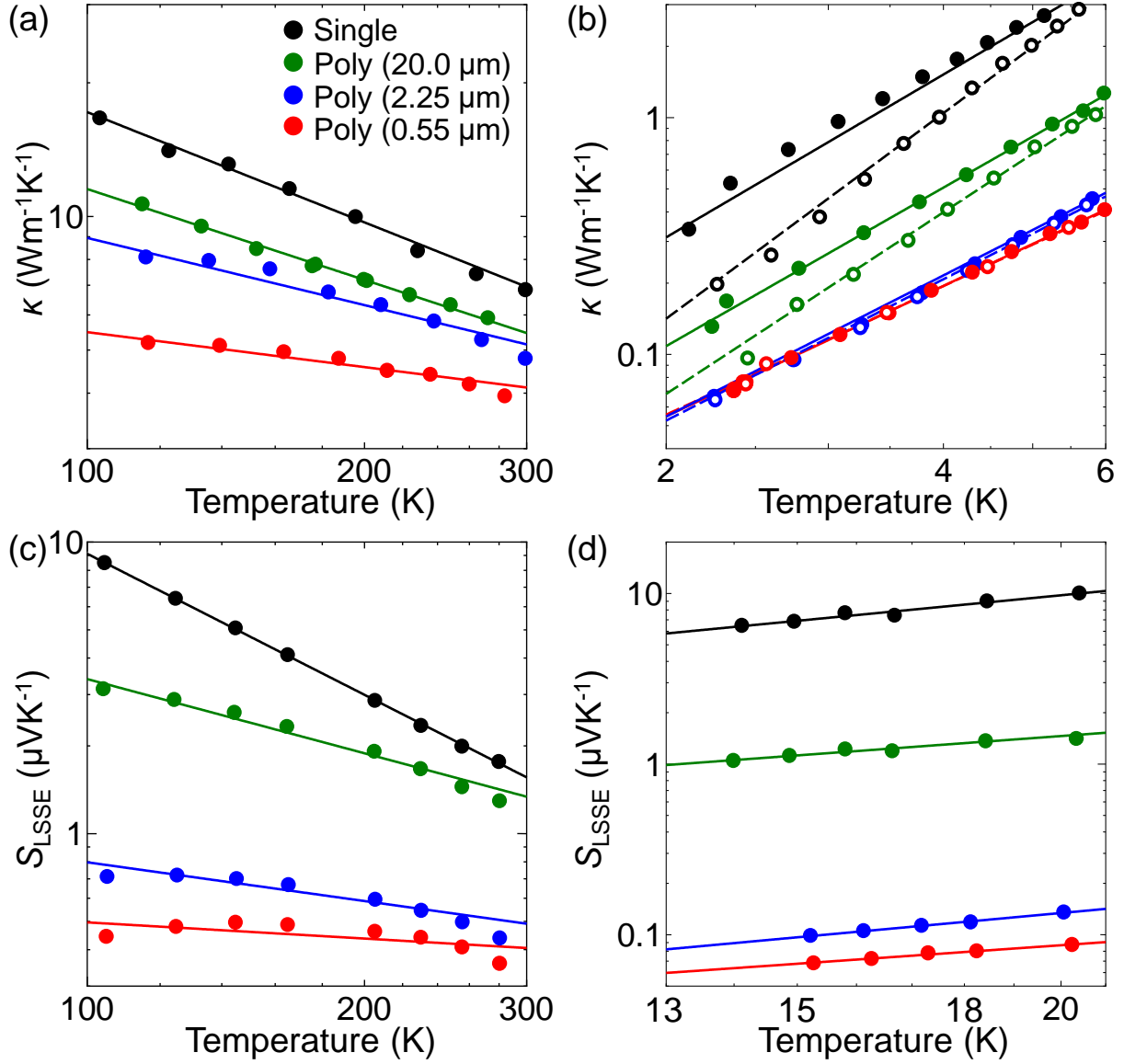

FIG. 5. Temperature dependence of thermal conductivity in single-crystalline YIG (black) and polycrystalline samples with average grain size of 20.0 μm (green), 2.25 μm (blue) and 0.55 μm (red) in a log-log scale in temperature range of (a) from 100 K to 300 K and (b) from 2 K to 6 K. Here solid and open circle represent thermal conductivity at $H = 0$ kOe and $H = 90$ kOe respectively. Dot lines represent the curves determined by fitting the measured thermal conductivity and thermopower using $CT^n$. (c),(d) Temperature dependence of LSSE

thermopower at $H = 1.8$ kOe in a log-log scale in temperature range of (c) from 100 K to 300 K and (d) from 13 K to 20 K.

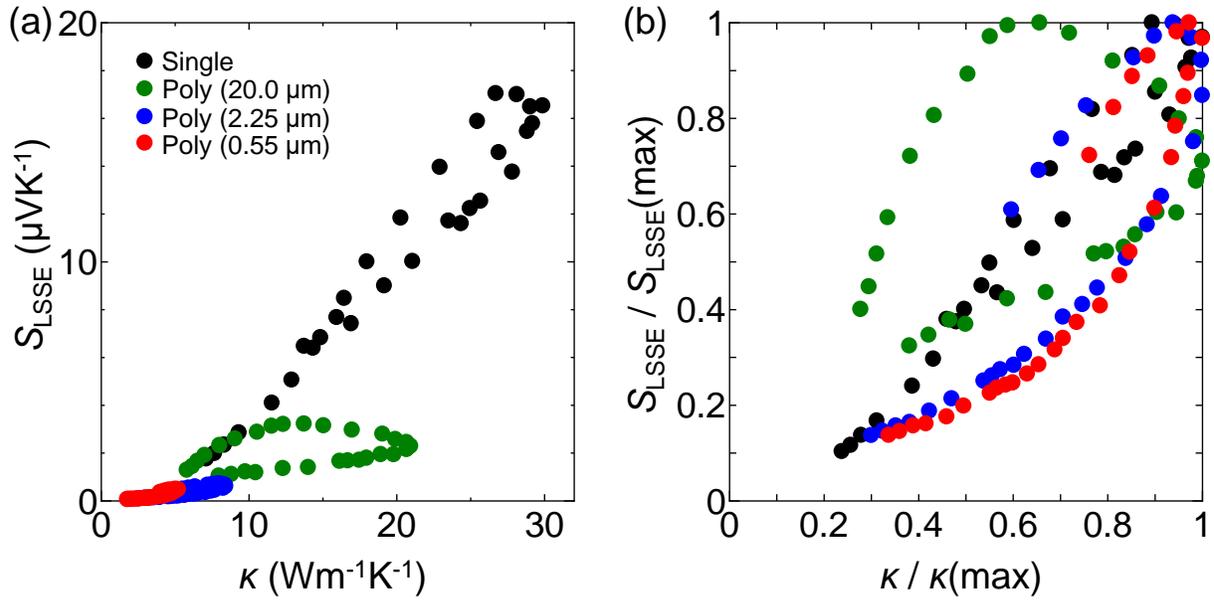

FIG. 6. (a) The correlation between thermal conductivity and thermopower of single-crystalline YIG (black) and polycrystalline samples with average grain size of 20.0 μm (green), 2.25 μm (blue) and 0.55 μm (red). (b) The thermal conductivity and thermopower are normalized by each max value.

TABLE I. Dimensions of all the samples in thermal and LSSE measurement.

|  | Dimensions in thermal measurement (mm) | Dimensions in LSSE measurement (mm) |
|---|---|---|
| Single | 2.24×0.76×7.00 | 4.06×1.48×0.89 |
| Commercial poly | 1.65×1.00×14.6 | 4.09×1.49×0.90 |
| Poly-2.25 μm | 1.96×1.73×9.06 | 4.03×1.54×1.56 |
| Poly-0.55 μm | 1.79×0.93×9.80 | 4.10×1.42×0.91 |

TABLE II. Milling time of YIG powder before sintering, and relative density and average grain size of sinterd sample.

|  | Milling time (hours) | Relative density (%) | Grain size (μm) |
|---|---|---|---|
| Single | - | - | - |
| Commercial poly | - | - | 20.0 |
| Poly-2.25 μm | 6 | 96.1 | 2.25 |
| Poly-0.55 μm | 50 | 96.1 | 0.55 |

TABLE III. The exponent $n$ which is determined by fitting the measured thermal conductivity and thermopower with a power-law function of $T$, $CT^n$, where $C$ and $n$ are fitting parameters and $T$ is temperature. The value inside the bracket is determined from the temperature dependence of the thermal conductivity at $H = 90$ kOe. The last column denotes the values of saturation magnetization $M$ at 300 K.

|  | $n$ |  |  |  | $M$ (emu/g) |
|---|---|---|---|---|---|
|  | $\kappa$ in high $T$ | $\kappa$ in low $T$ | $S_{\text{LSSE}}$ in high $T$ | $S_{\text{LSSE}}$ in low $T$ |  |
| Single | -0.82 | 2.28 (3.00) | -1.60 | 1.19 | 28.3 |
| Commercial poly | -0.68 | 2.26 (2.61) | -0.84 | 0.91 | 27.2 |
| Poly-2.25 μm | -0.50 | 2.07 (2.07) | -0.44 | 1.05 | 25.3 |
| Poly-0.55 μm | -0.26 | 1.86 (1.85) | -0.18 | 0.94 | 24.1 |

# AUTHOR INFORMATION

Corresponding Author

*E-mail: shiomi@photon.t.u-tokyo.ac.jp, UCHIDA.Kenichi@nims.go.jp.

## ACKNOWLEDGMENT

The thermal conductivity measurement was performed using facilities of the Cryogenic Research Center, the University of Tokyo. Structural characterization using EBSD was conducted at Advanced Characterization Nanotechnology Platform of the University of Tokyo supported by "Nanotechnology Platform" of the Ministry of Education, Culture, Sports, Science and Technology (MEXT), Japan. We thank S. Ito from Analytical Research Core for Advanced Materials, Institute for Materials Research, Tohoku University, for performing transmission electron microscopy on our samples. This work is partially supported by CREST "Scientific Innovation for Energy Harvesting Technology" (No. JPMJCR16Q5), PRESTO "Phase Interfaces for Highly Efficient Energy Utilization" (No. JPMJPR12C1) and ERATO "Spin Quantum Rectification Project" from JST, Japan, Grant-in-Aid for Scientific Research (B) (No. JP16H04274) and (A) (No. JP15H02012), Grant-in-Aid for Scientific Research on Innovative Area "Nano Spin Conversion Science" (No. JP26103005), from JSPS KAKENHI, Japan, NEC Corporation, the Noguchi Institute, and E-IMR, Tohoku University. A.M. and T.K. are supported by JSPS through a research fellowship for young scientists (No. JP16J09152 for A.M. and No. JP15J08026 for T.K.).